\documentstyle[amssymb]{article}

\newtheorem{proposition}{Proposition}
\newtheorem{lemma}{Lemma}
\newtheorem{corollary}{Corollary}

\begin{document}

\begin{center}

{\bf\Large Constrained Reductions of 2D dispersionless Toda Hierarchy, Hamiltonian Structure and Interface Dynamics.}\\*[3ex]

{\bf\Large J. Harnad$^1$, I. Loutsenko$^2$, O.Yermolayeva$^3$}\\*[3ex]

$^1$CRM, Universite de Montreal, C.P. 6128, succ. centre-ville, Montreal, Quebec, H3C 3J7, Canada.\\
e-mail: harnad@crm.umonreal.ca\\

\vspace{5mm}

$^2$Institute of Mathematics, University of Oxford, 24 - 29 St. Gilles', Oxford, OX1 3LB, England

e-mail: loutsenk@maths.ox.ac.uk

\vspace{5mm}

$^3$SISSA, Via Beirut 2-4, 34014, Trieste, Italy, e-mail: yermola@fm.sissa.it\\

\vspace{5mm}

\bf Abstract \end{center}

\begin{quote}

Finite-dimensional reductions of the 2D dispersionless Toda
hierarchy constrained by the ``string equation'' are studied.
These include solutions determined by polynomial, rational or
logarithmic functions, which are of interest in relation to the
``Laplacian growth'' or Hele-Shaw problem governing interface dynamics. The
consistency of such reductions is proved, and  the Hamiltonian
structure of the reduced dynamics is derived. The Poisson
structure of the rationally reduced dispersionless Toda
hierarchies is also derived.

\end{quote}

\begin{section}{Introduction.}

This paper concerns rational and logarithmic
reductions of the 2D dispersionless Toda hierarchy of integrable
equations (henceforth 2dToda). The subject is motivated by
important applications to problems in interface dynamics and
statistical physics.

Laplacian growth is a process that governs the dynamics of the
boundary in the plane separating two disjoint, open regions ${\mathcal D}_+$ and ${\mathcal D}_-$ in
which harmonic (scalar) fields are defined. These may be
interpreted as the pressure fields for two incompressible viscous
fluids (Hele-Shaw problem). The movement of the boundary is determined (according to
Darcy's law, in the case of viscous fluids) by equating the
normal velocity of the boundary to the boundary value of
the gradient of the field. In particular, one region (say, the
``interior'' region ${\mathcal D}_+$) may be chosen to be bounded and have constant
harmonic field (corresponding to zero viscosity) with the boundary
condition for the ``exterior'' ${\mathcal D}_-$ region at infinity to be such that
there is a unit sink, implying that the area of the interior
region grows linearly in time \cite{R}. Denoting the harmonic
field (e.g. the pressure) in the exterior  region by $P(X,Y)$,
this satisfies the conditions:
\begin{equation}
\begin{array}{r}
\Delta P(X,Y)=0\\
 P\to (4\pi)^{-1}\ln(X^2+Y^2) \quad {\rm as} \quad X^2+Y^2 \to \infty
\end{array}
\label{laplace}
\end{equation}
in the cartesian coordinates $(X,Y)$. The normalized exterior normal velocity at the boundary is given by:
\begin{equation} \label{normal}
\upsilon_{n}=-n\nabla P .
\end{equation}
where $n$ denotes an outward normal to $\partial {\mathcal D}_+$
In the case where the boundary is an analytic curve it is usual to use the Riemann mapping theorem
to introduce a time-dependent conformal map from the exterior of
the unit circle in the complex $w$ plane
to the exterior region ${\mathcal D}_-$ in the ``physical'' plane $z=X+iY$ taking the unit circle to the boundary $\partial {\mathcal D}_+$.
\begin{equation}
z=z(w,x), \quad w=\exp(i\phi), \quad 0<\phi<2\pi,
\label{un}
\end{equation}
where $x$ stands for the physical time of the Hele-Shaw problem. We choose this unusual notation for time for consistency with that used in the literature on the dispersionless integrable systems. 

Simple considerations \cite{PK}, \cite{Dyn}, \cite{MWZ} show (see Appendix A) that equations (\ref{laplace}, \ref{normal}) are  equivalent to
the following equation
\begin{equation}
{\rm Im}\left(\frac{\partial z}{\partial\phi}\frac{\partial\bar
z}{\partial x}\right)=w\left(\frac{\partial z(w,x)}{\partial
w}\frac{\partial \bar z(1/w,x)}{\partial x}-\frac{\partial
z(w,x)}{\partial x}\frac{\partial\bar z(1/w,x)}{\partial
w}\right)=1, \label{inter} \end{equation} 
where bar stands for
complex conjugation (and $\bar w=w^{-1}$ on the boundary curve). In our notations $\bar z(w)=\sum_i \bar z_i w^i$ if $z(w)=\sum_i z_i w^i$, while $\overline {z(w)}=\sum_i \bar z_i \bar w^i$.

Known as the Galin-Polubarinova equation in the Hele-Shaw problem,  eq (\ref{inter}) plays an essential role  in the
theory of infinite-dimensional integrable hierarchies in the dispersionless limit. The
relation between the boundary dynamics above and the dispersionless
limit of the integrable Toda hierarchy constrained by
(\ref{inter}) was shown in \cite{MWZ}.

Equation (\ref{inter}) may be interpreted as a constraint on an
infinite commuting set of dynamical systems defined in the space
${z(x,w)}$ of one-parameter families of conformal maps. This
constraint represents fixed points of an ``additional
symmetry'' \cite{OS} and is called the ``string equation'' in the theory of integrable systems.
 The most interesting aspect of such constrained 2dToda flows is that they admit
finite-dimensional reductions, which include so-called ``multi-finger''
solutions \cite{H}.
These solutions are of great importance in practical applications
and describe numerous phenomena, such as  viscous fingering in a
Hele-Shaw cell  \cite{H},\cite{M},\cite{ST} and pattern formation in
the quantum Hall effect \cite{ABWZ}.

In what follows, we consider finite-dimensional reductions of
(\ref{inter}) in the context of the 2dToda hierarchy. We first study
algebraic solutions of the problem, ignoring the real
structure and treating $z$, $\bar z$ as independent functions, and
$w$ as a formal variable. Returning to the applications to interface dynamics we identify bar with complex conjugation.
\end{section}

\begin{section}{2dToda hierarchy and string equation.}

The 2dToda hierarchy is a dispersionless limit \cite{TT} of the two-dimensional Toda hierarchy and is defined in terms of two functions $z(w,x)$
and $\bar z(w^{-1},x) $ of the form:
\begin{equation} \label{z}
z(w,x) = r(x)w + \sum\limits_{k=0}^{\infty} u_k(x) w^{-k},
\end{equation}
\begin{equation} \label{z-bar}
\bar{z}(w^{-1},x) = r(x)w^{-1} + \sum\limits_{k=0}^{\infty} \bar
{u}_k(x) w^k.
\end{equation}
The 2dToda flow equations are
\begin{equation}\begin{array}{ll}
\partial_{t_k} z = \{H_k , z\} & \qquad \partial_{\bar{t}_k} \bar{z} = \{\bar{H}_k , \bar{z}\}  \\
\partial_{t,_k} \bar{z} = \{H_k , \bar{z}\} & \qquad \partial_{\bar{t}_k} z = \{\bar{H}_k , z\}
\end{array}\label{zz}
\end{equation}
where the ''Poisson-Lax" bracket notation here denotes
\begin{equation} \label{Lax bra}
\{f,g\} := w \frac{\partial f}{\partial w} \frac{\partial g}{\partial x} - w \frac{\partial f}{\partial x} \frac{\partial g}{\partial w}
\end{equation}
and the coefficients $r(x), u_{k}(x), \bar u_{k}(x)$ are viewed
as coordinate functions on the phase space. The evolution functions are defined as follows
\begin{equation}
H_k=(z^k)_++1/2(z^k)_0, \quad \bar H_k=(\bar z^k)_-+1/2(\bar
z^k)_0, \label{dflows}
\end{equation}
where subscripts $\pm, 0$ denote the negative/positive and zero
parts of the formal Laurent expansion in $w$ (i.e. $f_+=\sum_{i>0}f_iw^i, \quad f_-=\sum_{i<0}f_iw^i $ if $f=\sum_if_iw^i$).

It is important to note that, despite the notation, (\ref{Lax bra}) is not really a Poisson bracket defined on the infinite-dimensional phase space
of the 2dToda system (\ref{zz}) (with coordinates $r,u_k,\bar u_k$), but rather a ``quasiclassical" $\hbar\to0$
limit of commutators in the Lax representation of the dispersive version of (\ref{zz})
(where $\hbar$ stands for the lattice spacing, see \cite{TT}). Values (\ref{dflows}) are dispersionless analogs of
the upper or lower diagonal parts of the powers of Lax matrices.

The 2dToda system is nevertheless an integrable Hamiltonian system of PDEs for the functions $r(x,t_1,...,\bar t_1 ...)$, $u_k(x,t_1,...,\bar t_1 ...)$, $\bar u_k(x,t_1,...,\bar t_1 ...)$, obtained by equating coefficients of (\ref{zz}) as Laurent polynomials in the dummy variable $w$. One can easily verify that vector fields (\ref{zz}) commute, i.e
$$
\frac{\partial^2}{\partial t_i \partial t_j}\left(\begin{array}{c} z \\ \bar z \end{array}\right)=\frac{\partial^2}{\partial t_j \partial t_i}\left(\begin{array}{c} z \\ \bar z \end{array}\right), \quad \frac{\partial^2}{\partial \bar t_i \partial t_j}\left(\begin{array}{c} z \\ \bar z \end{array}\right)=\frac{\partial^2}{\partial t_j \partial \bar t_i}\left(\begin{array}{c} z \\ \bar z \end{array}\right), \quad \frac{\partial^2}{\partial \bar t_j \partial \bar t_i}\left(\begin{array}{c} z \\ \bar z \end{array}\right)=\frac{\partial^2}{\partial \bar t_i \partial \bar t_j}\left(\begin{array}{c} z \\ \bar z \end{array}\right)
$$
due to the ''zero curvature" conditions
$$
\frac{\partial H_i}{\partial t_j}-\frac{\partial H_j}{\partial t_i}-\{H_i,H_j\}=0, \quad \frac{\partial H_i}{\partial \bar t_j}-\frac{\partial \bar H_j}{\partial t_i}-\{H_i,\bar H_j\}=0, \quad \frac{\partial \bar H_i}{\partial \bar t_j}-\frac{\partial \bar H_j}{\partial \bar t_i}-\{\bar H_i,\bar H_j\}=0,
$$
which follows from equations of motion (\ref{zz}) and definitions (\ref{Lax bra}),(\ref{dflows}) .

The Galin-Polubarinova equation (\ref{inter}) for the Hele-Shaw problem written in terms of the Poisson-Lax bracket (\ref{Lax bra})
\begin{equation} \label{string}
\{z(w,x), \bar{z}(w^{-1},x)\} = 1
\end{equation}
is the string equation.

It is fundamental that equation (\ref{string}) is invariant under the flows generated by (\ref{zz}). Indeed
\begin{equation}
\partial_{t_k}\left\{ z,\bar z\right\}=\left\{\partial_{t_k}{z},\bar z\right\}+\left\{z,\partial_{t_k}{\bar z}\right\}=
\left\{\{H_k,z\},\bar z\right\}+\left\{z,\{H_k,\bar z\}\right\}=\left\{\{ \bar z , z \} , H_k \right\}=0 ,
\label{invariance}
\end{equation}
where we have used (\ref{zz}), (\ref{string}) and the Jacobi identity for the Lax-Poisson bracket (\ref{Lax bra}).

Thus the string equation (\ref{string}) defines an invariant manifold under the Toda flows (\ref{zz}) and a reduction of
the Toda hierarchy. On the other hand, the Toda flows may be viewed as symmetries generating new solutions of Eq (\ref{string}).

\end{section}

\begin{section}{Reductions of the 2dToda hierarchy constrained by the string equation.}

The reduction of the 2dToda hierarchy by the string equation is
still a compatible set of infinite-dimensional dynamical
systems.

Indeed, as seen from (\ref{z})-(\ref{z-bar}) the string equation (\ref{string}) is a system of ODE's of the form
\begin{equation}
dr/dx=R(r,u_1,...,\bar u_1,...), \quad d u_i/dx=U_i(r,u_1,...,\bar u_1,...), \quad d \bar u_i/dx=\bar U_i(r,u_1,...,\bar u_1,...)
\label{ode}
\end{equation}

In what follows we will be interested in further ``functional"
reductions where $z, \bar z$ are polynomial, rational or
logarithmic functions of $w$. As  shown below, such reductions are
consistent with (\ref{zz}) (i.e. the corresponding ansatz for $z$
is preserved by the 2D Toda flows) if the string equation
(\ref{string}) holds. Thus, for consistency we need a double
(``functional" plus ``string") reduction. These pairs of reductions
define finite-dimensional invariant sub-manifolds in the phase
space of the general 2dToda hierarchy. Indeed, functional reduction
leaves a finite number of discrete indices in the ansatz for $z$
as a function of $w$, while the string equation fixes the dependence of $z$ on  $x$, leaving finite
number of degrees of freedom. These degrees of freedom are
connected with the integration constants of system (\ref{ode}), which
becomes finite-dimensional after functional reductions.

\begin{subsection}{Polynomial reductions.}

We begin with polynomial reductions of the 2d Toda hierarchy
\begin{equation}
z(w)=rw+\sum_{i=-N}^0u_iw^i
\label{poly1}
\end{equation}
\begin{equation}
\bar z(w^{-1})=rw^{-1}+\sum_{i=0}^N\bar u_iw^i
\label{poly2}
\end{equation}
constrained by the string equation (\ref{string}).

The following proposition states the consistency of the polynomial
reductions under the Toda flows:
\begin{proposition}\label{laurent}

If the string equation (\ref{string}) holds, then (\ref{poly1}),
(\ref{poly2}) is a finite dimensional subspace (of the phase space (\ref{z}), (\ref{z-bar})) invariant under the $t_i$ and $\bar t_i$ 2dToda
flows,
$0<i<N+2$ (\ref{zz}). This subspace  has dimension $2N+3$ with local coordinates
chosen as the initial values of the solutions $r=r(x)$, $u_{i}=u_{i}(x)$, $\bar u_{i}=\bar u_{i}(x)$ to the reduced string equation (\ref{ode}).

\end{proposition}

{\it Proof}: We must prove that $z$ remains of the form
(\ref{poly1}), under the flows generated by $H_k, \bar H_k$
provided the string equation (\ref{string}) holds. In other words,
we have to show that the evolution does not change the highest and
lowest degrees of the Laurent polynomial (\ref{poly1}). The proof
for $\bar z$ is analogous.

1. First we proceed with evolution of $z$ generated by $H_k$. The lowest degree of $w$ in $z$ (\ref{poly1}) is $-N$. Since
\begin{equation}
H_k=(z^k)_{>0}+1/2(z^k)_0=h_k(x)w^k+h_{k-1}(x)w^{k-1}+\dots+h_0(x)
\label{Ham}
\end{equation}
is a Laurent polynomial of positive degree in $w$,
the lowest degree of the bracket $\{z,H_k\}$ is not less than $-N$, as follows from (\ref{Lax bra}). On the other hand, the complement $z^k-H_k$ of (\ref{Ham}) is a polynomial in $1/w$ and so,
$$
\{H_k,z\}=-\{z^k-H_k,z\}
$$
is a Laurent polynomial whose highest degree in $w$ does not exceed $1$.

2. Unlike the $H_k$-flows, the form-invariance of $z$ (\ref{poly1}) under
the flows generated by $\bar H_k$ requires extra restrictions
on derivatives $\partial_{x} u_i, \partial_{x} \bar u_i$ (the string equation).

Since
$$
\bar H_k=(\bar z^k)_-+1/2(\bar z^k)_0
$$
is a Laurent polynomial of non-positive degree, it conserves the highest degree of
$z(w)$ under the evolution (\ref{zz}), but not the lowest degree, in general. Indeed,
$$
\bar H_k=\bar z^k-((\bar z^k)_++1/2(\bar z^k)_0)
$$
is a difference between $\bar z^k$ and a polynomial of nonnegative degree, so that
$$
\{\bar H_k, z\}=\{\bar z^k,z\}-\{((\bar z^k)_++1/2(\bar z^k)_0),z\}
$$
The second bracket in the last expression preserves the lowest degree of $z$. Thus the lowest degree of $ \{\bar H_k, z\}$ equals the lowest degree of $ \{\bar z^k,z\}=k\bar z^{k-1}\{\bar z, z\}$.
Imposing the extra restriction (\ref{string}) (string equation),
we see that the lowest degree does not exceed that of $\bar
z^{k-1}$. Since the lowest degree of $\bar z$ is $-1$ and that
of $z$ is $-N$, it follows that $k<N+2$.

Therefore, the form of (\ref{poly1}) is preserved by $t_i,\bar t_i$ Toda
flows with $i=1,..,N+1$, provided the string equation (\ref{string}) holds. This completes the proof.

It is easy to prove the converse to Proposition \ref{laurent}, i.e.
\begin{proposition}\label{converse} 
The string equation (\ref{string}) is a necessary condition for the existence of integrable polynomial flows of the form (\ref{dflows}), (\ref{poly1}), (\ref{poly2})
\end{proposition}

Proof: This essentially uses the same arguments as those leading to the result of Proposition \ref{laurent}. For the first $N+1$ $z$ flows to be consistent with the polynomial reduction it is necessary that the Laurent expansion of $\{\bar z,z\}$ 
be of the form $\{\bar z,z\}=f(x)+\sum_{i>0}U_i(x)/w^i$. The corresponding condition for the $\bar z$ flows is  $\{\bar z,z\}=\bar f(x)+\sum_{i>0}\bar U_i(x)w^i$. Both can be simultaneously satisfied only if
$$
\{\bar z,z\}=f(x)=\bar f(x)
$$
is independent of $w$.
Differentiating with respect to $t_i$ (or $\bar t_i$), i.e. along the flow lines, using the equations of motion (\ref{zz}), and the Jacobi identity we get
$$
\frac{\partial f}{\partial t_i}=\{H_i,f\}=w\frac{\partial f}{\partial x}\frac{\partial H_i}{\partial w}.
$$
Since $w\frac{\partial H_i}{\partial w}$ is $w$-dependent, while $f$ is not, it follows that $\frac{\partial f}{\partial x}=\frac{\partial f}{\partial t_i}=0$ and similarly $\frac{\partial f}{\partial \bar t_i}=0$. Therefore $f$ is a constant in $x$ and also constant along the flows, implying (up to a constant scaling) that the string equation (\ref{string}) holds.  

In the following subsections we consider rational and logarithmic reductions as well as their Hamiltonian structures. Polynomial
solutions can then be viewed as a special case.
Nevertheless the corresponding limiting procedures are rather cumbersome and it is easier to
consider the polynomial case separately. A derivation of the canonical linearizing variables in the polynomial case
is given in Appendix B. This closely follows the analogous procedure given in Section \ref{poisson} for the logarithmic case.
Local canonical lianearizing variables, which are constant and linear in the hierarchy times, turn out to be ``Richardson's harmonic moments" of the Hele-Shaw (Laplacian growth) problem when
$z(w)$ and $\bar z(\bar w)$ are identified with the conformal mapping from the $w$ to the $z$ plane. This is a simple proof
of the relation (originally established in \cite{MWZ}) between the ``times" of the 2dToda hierarchy and harmonic moments of the exterior Hele-Shaw problem.

An important consequence of the present section is the fact that the string equation (\ref{string}) necessarily holds for reduced conformal mappings evolving under a sufficient number of 2dToda flows (i.e. sufficient for complete integrability). In the
context of interface dynamics, {\it Darcy's law thus turns out to be a corollary of integrability} for reduced systems.

\end{subsection}

\begin{subsection}{Rational reductions.}

We now consider a rational reduction for $z(w)$ and $\bar z(w)$, $N>1$ given by
\begin{equation} \label{rat1z}
z(w)=\frac {q_{N+1}(w)} {p_N(w)} = \frac {rw^{N+1}+\sum_{i=0}^{N} a_i
w^i } { w^N + \sum_{i=0}^{N-1} b_i w^i  },  \end{equation}
\begin{equation} \label{rat1zz}
\bar z(w^{-1})=\frac {\bar q_{N+1}(w^{-1})} {\bar p_{N}(w^{-1})} = \frac
{rw^{-(N+1)}+\sum_{i=0}^{N} \bar a_i w^{-i} } { w^{-N} +
\sum_{i=0}^{N-1} \bar b_i w^{-i} } .  \end{equation}
The following Lemma states the consistency such rational reductions under the
$t_i$ Toda flows for (\ref{rat1z}) and the $\bar t_i$ flows for (\ref{rat1zz}).

\begin{lemma}\label {Rat-half-flows}
The form of the function $z(w)$ in (\ref{rat1z}) is invariant under the $t_i$ flows and the form (\ref{rat1zz}) of $\bar z(w^{-1})$ is invariant under the $\bar t_i$ flows for any $i>0$.
\end{lemma}
{\it Proof}:
Consider the flows generated by $H_k$.
Since
\begin{equation}
H_k=(z^k)_++1/2(z^k)_0
\label{one}
\end{equation}
is a polynomial of nonnegative degree in $w$, its complement $z^k-H_k$ is a polynomial in $1/w$.
The Laurent expansion of brackets (\ref{Lax bra}) $ \{H_k,z\}=-\{z^k-H_k,z\} $ around infinity therefore has the following form
\begin{equation}
\{H_k,z\}=k_1(T)w+k_0(T)+k_{-1}(T)w^{-1}+... \quad .
\label{lau}
\end{equation}
However the Lax bracket (\ref{Lax bra}) also implies that it is  rational of the form
$$
\{H_k,z\}=Q(w)/p_N(w)^2 .
$$
It follows from (\ref{lau}) that the degree of polynomial $Q(w)$ does not exceed $2N+1$.

On the other hand,
$$
\partial_{t_k} z=P(w)/{p_N}^2(w), \quad P(w)=p_N\partial_{t_k}q_{N+1}-q_{N+1}\partial_{t_k}p_N,
$$
where the degree of the polynomial $P(w)$ also does not exceed $2N+1$.
Equating the coefficients of polynomial $P(w)-Q(w)$ to zero, we get a system
of differential equations for $r,a,b$. The number of equations is $2N+2$.
Thus we get a compatible system of differential equations for $2N+2$
unknowns $r,a,b$. A similar argument shows the form invariance of (\ref{rat1zz}) under the $\bar t_k$ flows.

The consistency of the rational reduction defined in (\ref{rat1zz}) under the $t_i$ flows and (\ref{rat1z}) under $\bar t_i$ flows, respectively requires additional restrictions such as the string equation. But this only suffices to ensure form invariance under $t_1$ and $\bar t_1$ flows respectively.

\begin{lemma}\label{Rat}

The string equation (\ref{string}) is a sufficient condition for $z(w)$ and $\bar z(w)$ ((\ref{rat1z}),
(\ref{rat1zz})) to be form invariant under the first two-Toda flows. Flows generated by $H_k$, $\bar H_k$ with $k>1$ are inconsistent with the rational reduction (\ref{rat1z}), (\ref{rat1zz}).

\end{lemma}

Remark: In the next section, we will show how to choose $4N+2$ independent commuting flows preserving the rational form of $z$ (\ref{rat1z}) and $\bar z$ (\ref{rat1zz}), generated by vector fields that are infinite linear combinations of those generating the 2dToda hierarchy.\\

{\it Proof}: 
In order to prove that $z$ remains of the same form (\ref{rat1z}) under
the $\bar t_i$ Toda flows
we would have to find a compatibility condition for a system of differential equations for $r(T)$, $a(T)$, $b(T)$, induced by (\ref{zz}).

Consider evolution of $z$ along the flows generated by $\bar H_k$. Again we write
$$
\partial z / \partial{\bar t_k} =S(w)/{p_N}^2(w), \quad S(w)=p_N\partial_{\bar t_k}q_{N+1}-q_{N+1}\partial_{\bar t_k}p_N
$$
with $S(w)$ a polynomial in $w$ of order at most $2N+1$.

Since
\begin{equation}
\bar H_k=(\bar z^k)_-+1/2(\bar z^k)_0 = \bar h_k w^{-k} + \bar h_{k-1} w^{-k+1} + \dots + \bar h_{0}
\label{two}
\end{equation}
is a Laurent polynomial of non-positive degree in $w$, the Laurent expansion of the corresponding bracket has the following form
$$
\{\bar H_k,z\}=k_1(T)w+k_0(T)+k_{-1}(T)w^{-1}+...
$$
The definition of the Lax bracket (\ref{Lax bra}) and the expansion (\ref{two}) implies that
$$
\{\bar H_k,z\}=(U(w)+R(1/w))/p_N(w)^2
$$
where $U(w)$ is a polynomial of degree at most $2N+1$ and $R=w^{-1}(\eta_{k-1}(T)w^{-k+1}+\dots+\eta_0)$ is a polynomial in $w^{-1}$. If $R(1/w)=0$, the number of equations will not exceed the number of unknowns. However, as we show below, the string equation
does not imply the vanishing of $R$.

Introduce the new variable $y=w^{-1}$. The vanishing of $R$ is equivalent to the following form of expansion in $y$
\begin{equation}
\{\bar H_k,z\}=f_0(T)+f_{-1}(T)y^{-1}+... \quad .
\label{exp}
\end{equation}
Consider this as a function of $y$ and, as before, express $\bar H_k$ as a difference of $\bar z_k$ and a polynomial of
nonegative degree. Then
\begin{eqnarray} \label {mixed}
\{\bar H_k, z\}=\{\bar z^k,z\}-\{((\bar z^k)_++1/2(\bar z^k)_0),z\}= \nonumber\\
= k \bar z^{k-1}\{ z,\bar z \} -\{((\bar z^k)_++1/2(\bar z^k)_0),z\}
\end{eqnarray}
Imposing the string equation (\ref{string}) we see that the only flow that allows the above expansion (\ref{exp})
corresponds to $k=1$.
A similar argument shows that only the $t_1$ flow preserves the form (\ref{rat1zz}) of $\bar z$ even if the string equation is imposed.

As seen from the above proof, there are only two flows generated by evolution operators of the form (\ref{dflows}) compatible with the rational reduction. In fact, we
should not expect
more invariant flows associated to the simple poles at $w=0$ and $w=\infty$. In the polynomial case, the number
of invariant
flows was equal to the number of variables (polynomial coefficients), since one can associate $n$ invariant
flows to a pole of $n$ th order, and the poles at zero and infinity
are immovable.

However, below we introduce additional flows related to  movable singularities of $z(w)$ and $\bar z(w^{-1})$ which do preserve the rational reduction (\ref{rat1z}), (\ref{rat1zz}), extending results of Krichever for the KP case.

\end{subsection}

\begin{subsection}{Additional flows for rational reductions of dKP hierarchy.}

In this subsection we recall the theory of flows related to poles at a finite number of finite points applying an approach previously used by Krichever for dKP hierarchy.
In the dispersionless limit, the dKP and 1dToda hierarchies are quite similar, while in 2dToda the existence of finite-dimensional
reductions requires extra constraints.

Let us start by recalling \cite {EOR}, that on the phase space of extended
Benney systems, i.e. rational dKP, there arise some new flows related to the pole
structure of the corresponding maps. These additional flows were
introduced by Krichever (see  \cite{Kri3}).

Consider the partial fraction expansion under the above assumptions and the flows of the $t_i$ type only. For the dKP hierarchy \cite{EOR}, \cite{T} these reductions have the form
\begin{equation} \label{z-add}
z(w)=w+u_0+\sum_{\alpha=1}^{N}\frac{u_{\alpha}}{w - w_\alpha} .
\end{equation}
The new flows associated to the poles are defined similarly to the polynomial case
\begin{equation}\label{add-half-flows}
\partial_{t_{k,\alpha}} z = \{B_{k,\alpha} , z\}, \quad \alpha= \infty, 1,2,... \quad k=0,1,2,...
\end{equation}
The evolution functions associated with the pole structure of $z$ are as follows :
$$
B_{k,\infty}= (z(w)^{k})_{\geq 0}, \quad ()_{\geq 0}:=()_++()_0
$$
for an immovable pole at infinity, while for  each finite-distance pole there appear additional flows
with evolution functions:
$$
B_{k,\alpha}= (z(w)^{k})_\alpha ,\quad B_{0,\alpha}=\log (w_\alpha-w) .
$$
Here $z(w)_\alpha$ denotes the negative part of a formal Laurent expansion of $z(w)$ near the pole $w_\alpha$:
\begin{equation}
f(w)_\alpha=\sum_{i>0} \frac{f_i}{(w-w_\alpha)^i}, \quad {\rm if} \quad f=\sum_{i \in Z} \frac{f_i}{(w-w_\alpha)^i} .
\label{singular}
\end{equation}
These additional flows commute amongst themselves and with the
ordinary 1dToda or dKP flows (associated with poles at infinity) .

\end{subsection}

\begin{subsection}{Additional invariant flows of 2dToda system.}

In the following, we will look at reductions of the 2dToda systems analogous to the above Benney-Krichever reductions of KP. We will thus choose the rational functions $z(w)$, $\bar z(1/w)$ appearing in (\ref{rat1z}) and (\ref{rat1zz}) to have only simple poles and express these in the partial fraction form
\begin{equation} \label {bar-z-add} \begin{array}{l}
z(w)=rw+u_0+\sum_{j=1}^n\frac{u_j}{w-w_j},\\
\bar z(1/w)=r/w+\bar u_0+\sum_{j=1}^n\frac{\bar u_j}{1/w -\bar w_j} .\end{array}
\end{equation}
Now, introduce a new set of $4n+2$ evolution functions $H_{0,j},H_{1,j},H_{1,\infty},\bar H_{0,j},\bar H_{1,j},\bar H_{1,\infty}$ defined by
\begin{equation} \label{add-ev-op}
H_{k,j}(w) = B_{k,j}(w) - \frac{1}{2}B_{k,j}(w=0) , \quad  \bar H_{k,j}(y) = \bar B_{k,j}(y) - \frac{1}{2} \bar B_{k,j}(y=0),
\end{equation}
where $k=0,1$, $j=1..n$ and
\begin{equation}
\label{singul}
\begin{array}{lll}
B_{1,\infty}(w)=(z(w))_{\geq 0}, & B_{0, j}=\log\left(r(w_j-w)\right), & B_{1, j}(w)=(z(w))_j \\
\bar B_{1,\infty}(y) = (\bar z(y))_{\geq 0} ,& \bar B_{0, j}=\log\left(r(\bar w_j-y)\right) , & \bar  B_{1,j}(y)= (\bar z(y))_j, \quad y=1/w.
\end{array}
\end{equation}
We denote the flow variables associated with $H_{0,j}$ as $\tau_{2j}$, those associated with $H_{1,j}$ as $\tau_{2j-1}$, and that associated with $H_{1,\infty}$ as $\tau_0$. The flow variables associated with $\bar H_{0,j}$, $\bar H_{1,j}$ and $\bar H_{0,\infty}$ are denoted $\bar\tau_{2j}, \bar\tau_{2j-1}, \bar\tau_{0}$ respectively. Correspondingly, we denote the evolution functions
\begin{equation}\begin{array}{ll}
h_0=H_{1,\infty}=rw+u_0/2, & \bar h_0=\bar H_{1,\infty}=r/w+\bar u_0/2 ,\\
h_{2j-1}=H_{1,j}=\frac{u_j}{w-w_j}+\frac{u_j}{2w_j}, & \bar h_{2j-1}=\bar H_{1 , j}=\frac{\bar u_j}{\bar w_j-1/w}+\frac{1}{2}\bar u_j/\bar w_j ,\\
h_{2j}=H_{0,j}=\log(w_j-w)+1/2\log(r/w_j), & \bar h_{2j}=\bar H _ {0,j} =\log(\bar w_j-1/w)+1/2\log(r/\bar w_j) .
\label{evolve}
\end{array}
\end{equation}
Then the following proposition holds

\begin {proposition} \label {poles-flows}
The $4n+2$ commuting Toda-Krichever flows
\begin{equation}
\begin{array}{ll}
\partial_{\tau_j} z =\left\{h_j,z\right\}, & \partial_{\bar \tau_j} z =\left\{\bar h_j,z\right\} \\
\partial_{\tau_j} \bar z =\left\{h_j,\bar z\right\}, & \partial_{\bar \tau_j} \bar z=\left\{\bar h_j,\bar z\right\} \\
\end{array}, \qquad j=0..2n
\label{eqs_poles}
\end{equation}
preserve the rational form of
$z(w)$ and  $\bar z(1/w)$ (\ref{bar-z-add}) (or equally (\ref{rat1zz}, \ref{rat1z})) provided the string equation (\ref{string}) holds. The dimension of the reduced phase space equals $4n+3$.
\end {proposition}
We defer the proof, since this follows as a limiting case of the more general logarithmic reduction introduced in the next subsection.

As in the polynomial case, the total number of form invariant flows preserving the string equation
equals the dimension of the reduced phase space minus one. In what
follows we show that these flows are Hamiltonian. Since the
dimension of the phase space is odd and equals $4n+3$, it is,
in fact a Poisson manifold whose symplectic leafs have
dimension $4n+2$, which is exactly the number of commuting Toda-Krichever
flows.

The above result holds for a more
general setting. Below we introduce a logarithmic reduction of the
2dToda hierarchy and prove an analog of proposition \ref{poles-flows} for logarithmic functions. Proposition 
\ref{poles-flows} follows as a limiting case.

\end{subsection}

\begin{subsection}{Logarithmic flows.}

It is easier to prove the consistency of the rational reductions with the dynamics of the 2dToda system by first considering
the more general logarithmic functions and then taking limits in which the branch points degenerate in pairs. Let us set
\begin{equation}
\begin{array}{l}
z=r(x)w+u(x)+\sum_{i=1}^{n+1}a_i\log(w_i(x)-w) , \\
\bar z=r(x)w^{-1}+\bar u(x)+\sum_{i=1}^{n+1}\bar a_i\log(\bar w_i(x)-w^{-1}) ,
\end{array} \label{zln}
\end{equation}
where $a_i, \bar a_i$ are arbitrary constants, subject to the conditions
\begin{equation}
\sum_{i=1}^{n+1}a_i=0,\quad \sum_{i=1}^{n+1}\bar a_i=0,
\label{suma}
\end{equation}
which ensure absence of logarithmic singularities at infinity.

Introduce evolution functions as follows:
\begin{equation}
\begin{array}{l}
{\mathcal H}_0=r(x)w+\frac{1}{2}u(x), \quad \bar {\mathcal H}_0=\bar r(x)w^{-1}+\frac{1}{2}\bar u(x) , \\
{\mathcal H}_j=\log(w_j(x)-w)+\frac{1}{2}\log(r(x)/w_j(x)) \\
\bar {\mathcal H}_j=\log(\bar w_j(x)-w^{-1})+\frac{1}{2}\log(r(x)/\bar w_j(x)) ,\end{array} j=1..n+1 .
\label{logh}
\end{equation}
We then have
\begin{proposition}\label{ln}
The string equation (\ref{string}) is a necessary and sufficient condition for the existence of $2n+4$ commuting flows on the $2n+5$ dimensional space of functions of the form (\ref{zln})
\begin{equation} 
\begin{array}{ll}
\partial_{\tau_i} z=\left\{{\mathcal H}_i ,z\right\}, & \partial_{\bar \tau_i} z=\left\{\bar {\mathcal H}_i,z\right\} \\
\partial_{\tau_i} \bar z=\left\{{\mathcal H}_i,\bar z\right\}, & \partial_{\bar \tau_i} \bar z=\left\{\bar {\mathcal H}_i, \bar z\right\}
\end{array}, i=0..n+1 ...n+1 \label{zzt}
\end{equation}
\end{proposition}
In other words, the flows generated by (\ref{zzt}), (\ref{logh}) are tangential to the manifold of such
logarithmic functions if the string condition is imposed (and conversely). We
therefore have $2n+4$ flows leaving invariant a $2n+5$ dimensional
sub-manifold of the 2dToda system. These will be shown to be infinite linear combinations of vector fields generating the 2dToda flows (\ref{zz}). Equations (\ref{ztolog}, \ref{logtot}) will show how to express corresponding 2dToda times $t_i,\bar t_i$ in terms of the logarithmic flow parameters $\tau_i,\bar \tau_i$. 

Proof

1.{\it Commutativity }: This will be shown to follow as Corollary \ref{logint} to Proposition \ref{int} below.

2.{ \it Consistency of  (\ref{zln}) with equations of motion (\ref{zzt})}: As in the polynomial and rational cases
the consistency of the $\partial_{\tau_i}z$ and $\partial_{\bar \tau_i}\bar z$ equations with the logarithmic reduction (\ref{zln}) follows from the fact that the structure of the Lax-Poisson brackets $\{{\mathcal H}_i,z\}$, $\{\bar{\mathcal H}_i, \bar z\}$ is identical to the infinitesimal deformations in $z$ and $\bar z$ induced by ones in the functions $ r, \bar u, u, \bar w_i, w_i$. However  the consistency of the $\partial_{\tau_i}\bar z$ and
$\partial_{\bar \tau_i}z$ equations with the logarithmic reduction
requires an extra constraint, the string equation. First, we prove this for $\partial_{\bar \tau_i}z$.

Differentiating $z$ in (\ref{zln}) with respect to $\bar\tau_i$ using the equation of motion (\ref{zzt}), we get
\begin{equation}
\partial_{\bar \tau_i}z=w\partial_{\bar \tau_i}r+\partial_{\bar \tau_i}u+\sum_{j=1}^{n+1}\frac{a_i \partial_{\bar \tau_i}w_j}{w-w_j}=\{\bar {\mathcal H}_i,z\} .
\label{ztaubar}
\end{equation}
The left-hand side of (\ref{ztaubar}) may contain any terms that are linear in $w$ at $w=\infty$ and simple poles at $w=w_j, j=1..n+1$. Since the Lax-Poisson bracket (\ref{Lax bra}) is a bi-derivation, the singularities in $w$ that may occur in $\{{\mathcal H}_i,z\}$ consist either of simple poles at $w=w_j, j=1..n+1$, linear terms at $w=\infty$ or a simple pole at $w=1/\bar w_i$. We show that the latter is absent.

Note that $\bar z$ can be represented as a sum over $\bar{\mathcal H}_i, i=0..n+1$ plus a $w$-independent function $f(x)$;
\begin{equation} \label{gau-z}
\bar z=\sum_{i=0}^{n+1}\bar a_i\bar {\mathcal H}_i+f(x),
\end{equation}
where $ \bar a_0:=1, \quad f(x)=\frac{1}{2}\left(\bar u(x)-\sum_{i=1}^{n+1}a_i\log\bar w_i(x)\right) $.
Now, using (\ref{gau-z}) we get
\begin{equation}
\bar a_i\{\bar {\mathcal H}_i, z\}=\{\bar z-\sum_{j\not=i}\bar a_j \bar {\mathcal H}_j-f(x),z\}=\{\bar z, z\}-\sum_{j\not=i}\bar
a_j\{\bar {\mathcal H}_j,z\}-\{f(x),z\} .
\label{ratprove} 
\end{equation}
Since $f(x)$ is independent of $w$, the last term in (\ref{ratprove}) contains only linear terms in $w$ plus simple poles at $w_j, j=1..n+1$. The term $\sum_{j\not=i}\bar a_j\{\bar {\mathcal H}_j,z\}$ could, in principle, contain poles at $w=1/\bar w_j$ for $j\not=i$, but these must cancel since $\{\bar {\mathcal H}_i,z\}$ contains no such poles. Furthermore, this term contains no pole at $w=1/\bar w_i$,
because $\bar {\mathcal H}_i$ is omitted in the sum.
Since the string equation (\ref{string}) holds by assumption, we see that the rhs of (\ref{ztaubar}) only contains terms of the type induced by infinitesimal deformations in the functions $r,u, w_j$, showing that the $\partial_{\bar \tau_i} z$ equation, together with the string equation, is compatible with the reduction (\ref{zln}). The proof for the $\partial_{\tau_i}\bar z$ equation is similar.

The above proves the sufficiency of the string equation for the validity of Proposition \ref{ln}. The necessity is proved similarly to Proposition \ref{converse}. By equation (\ref{ratprove}), if the rational reduction (\ref{zln}) is preserved under the $\bar \tau_i$ flows, $\{\bar z, z\}$ can have no pole at $w=1/\bar w_i$ for any $i$ or at zero. Similarly to preserve (\ref{zln}) under the $\tau_i$ flows it can have no poles at $w_i, i=1..n+1$ or $\infty$. But since the only possible poles in $\{\bar z, z\}$ are at these points, we conclude that $\{\bar z, z\}$ is constant in $w$ and hence can only be a function (say $q$) of $x,\tau,\bar\tau$ 
$$
\{\bar z, z\}=q
$$
Differentiating the last equation with respect to $\tau_i$ (or $\bar\tau_i$) using the equation of motion (\ref{zzt}), the Jacobi identity and the definition of the Lax-Poisson bracket we get
$$
\frac{\partial q}{\partial \tau_i}=w\frac{\partial q}{\partial x}\frac{\partial {\mathcal H}_i}{\partial w}.
$$
Since $w\frac{\partial {\mathcal H}_i}{\partial w}$ is $w$-dependent, while $q$ is not, it follows that $ \frac{\partial q}{\partial x}=\frac{\partial q}{\partial \tau_i}=0$ (similarly $\frac{\partial q}{\partial\bar \tau_i}=0$). Therefore $q={\rm const}$ and the string equation (\ref{string}) holds.

The demonstration that the reduced phase space is of dimension $2n+5$ is similar to the polynomial case. We view the string equation (\ref{string}) together with the logarithmic reduction (\ref{zln}) as a set of $(2n+5)$ first order ODE's for the $(2n+5)$ functions $r(x), u(x), \bar u(x), w_i(x), \bar w_i(x), i=1..n+1$
$$
\frac{dr}{dx}=R(r,u,\bar u,w_1,...,\bar w_1,...), \quad \frac{du}{dx}=U(r,u,\bar u,w_1,...,\bar w_1,...), \quad \frac{d\bar u}{dx}=\bar U(r,u,\bar u,w_1,...,\bar w_1,...)
$$
$$
\quad \frac{d w_i}{dx}=W_i(r,u,\bar u,w_1,...,\bar w_1,...), \quad \frac{d \bar w_i}{dx}=\bar W_i(r,u,\bar u,w_1,...,\bar w_1,...), \quad i=1..n+1
$$
which, at least locally, determine the dependence of these functions uniquely in terms of their values $r_0, u_0, \bar u_0, w_{i0}, \bar w_{i0}, i=1..n+1$ at some initial value $x=x_0$
$$
r=r(r_0, u_0, \bar u_0, w_{1,0},...,\bar w_{1,0} ,...,x),\quad u=u(r_0, u_0, \bar u_0, w_{1,0},..., \bar w_{1,0},...,x),\quad \bar u=\bar u(r_0, u_0, \bar u_0, w_{1,0}.. \bar w_{1,0}...,x)
$$
$$
w_i=w_i(r_0, u_0, \bar u_0, w_{10}..., \bar w_{1,0}...,x),\quad \bar w_i=\bar w_i(r_0, u_0,\bar u_0, w_{1,0},..,\bar w_{1,0}...,x) , \quad i=1..n+1
$$ 
The $2n+5$ initial values $r_0, u_0, \bar u_0, w_{i0}, \bar w_{i0}, i=1..n+1$ may be viewed as coordinates of the reduced phase space.

Proof of Proposition \ref{poles-flows}: This follows from taking limits in which the logarithmic branch points $w_{2i}$, $w_{2i-1}$ coalesce in pairs. 

Setting
\begin{equation}\begin{array}{lll}
a_{2i-1}=1/\epsilon, & a_{2i}=-1/\epsilon & w_{2i}=w_{2i-1}+\epsilon u_i\\
\bar a_{2i-1}=1/\epsilon, & \bar a_{2i}=-1/\epsilon & \bar w_{2i}= \bar w_{2i-1}+\epsilon \bar u_i \end{array}
\label{ratlim} \end{equation}
in (\ref{zln}) we get (\ref{bar-z-add}) in the $\epsilon \to 0$ limit.

The evolution functions (\ref{evolve}) , generating flows on the space of rational reduction (\ref{bar-z-add}), are then obtained as follows:
$$\begin{array}{ll}
h_0=\lim_{\epsilon=0}{\mathcal H}_0, & \bar h_0=\lim_{\epsilon=0}\bar {\mathcal H}_0\\
h_{2i-1}=\lim_{\epsilon=0}\frac{1}{\epsilon}({\mathcal H}_{2i}-{\mathcal H}_{2i-1}),&
\bar h_{2i}=\lim_{\epsilon=0}\frac{1}{\epsilon}(\bar {\mathcal H}_{2i}-\bar {\mathcal H}_{2i-1})\\
h_{2i}=\lim_{\epsilon=0}{\mathcal H}_{2i-1},& \bar h_{2i-1}=\lim_{\epsilon=0}\bar {\mathcal H}_{2i-1} .
\end{array}
$$

We thus obtain evolution functions for the rational reductions, where $z$, $\bar z$ have simple poles only, as a limiting degenerating logarithmic case.
One may deduce structures related to general rational reductions by degenerating
arbitrary numbers of logarithmic singularities at different finite points as well as at infinity.

\end{subsection}

\end{section}

\begin{section}{Poisson structure of logarithmic reductions of the 2Toda hierarchy.}

In this section we study the Poisson structure of logarithmic
reductions of the 2dToda hierarchy, which will be shown to define finite-dimensional completely integrable
systems. We find explicit expressions leading to a canonical
Hamiltonian structure on the phase space of rational reductions of
the 2dToda system.

The Poisson structure of rational reductions of the 1dToda hierarchy (which are infinite-dimensional, having no string equation constraint) is described in Appendix C. It is related to the Poisson structure of the Benney system considered in \cite{Pavlov}

\begin{subsection}{Hamiltonians, Action-Angle variables.}

Let us now introduce the following $2n+5$ functions on the reduced phase space defined by (\ref{zln}) with $z, \bar z$ solutions to (\ref{string}), extended by the auxiliary variable $x$
\begin{equation}
r,u,\bar u, w_1, ..., w_{n+1}, \bar w_1,...\bar w_{n+1} \to Q, I_0, I_1, ... I_{n+1}, \bar I_0, \bar I_1, ... \bar I_{n+1},
\label{change1} 
\end{equation}
where the new variables are defined as follows
$$\begin{array}{l}
I_0=\frac{1}{2i\pi}\oint_{\infty}\bar z d\ln z, \quad \bar I _0 =  \frac{ 1 } { 2i \pi   } \oint _{0} z d \ln \bar z, \\
I_j=\frac{1}{2 i \pi}\oint_{w_j}\bar z dz, \quad \bar I_j=\frac{1}{2i \pi}\oint_{1/\bar w_j} z d\bar z, \quad j=1..n+1 , \\
Q=\frac{1}{4 i\pi}\sum_{i=0}^{n+1} \oint_{1/{\bar w_i}} z d \bar z + \oint_{w_i} \bar z d z .\end{array}
$$
Using (\ref{zln}), we may evaluate the contour integrals in the last equation, expressing these explicitly in terms of the old parametrization $r,u,\bar u,w_j,\bar w_j, j=1..n+1$
\begin{equation}
\begin{array}{l}
I_0=\bar z(1/w=0)=\bar u+\sum_{j=1}^{n+1}\bar a_j\ln(\bar w_j), \quad \bar I_0=z(0)= u + \sum_{j=1}^{n+1} a_j\ln(w_j), \\
I_j=a_j\bar z(w_j^{-1})=a_j\left(rw_j^{-1}+\bar u+\sum_{k=1}^{n+1}\bar a_k\ln(\bar w_k-w_j^{-1})\right), \\
\bar I_j=\bar a_j z(\bar w_j^{-1})=\bar a_j\left(r\bar w_j^{-1}+u+\sum_{k=1}^{n+1}a_k\ln(w_k-\bar w_j^{-1})\right), \quad j=1..n+1, \end{array} \label{inte}
\end{equation}
\begin{equation}
\begin{array}{l}
Q=\frac{1}{2}r\left(\left(\frac{\partial z(w)}{\partial w}\right)_{w=0}+\left(\frac{\partial \bar z(1/w)}{\partial (1/w)}\right)_{1/w=0}\right)-\frac{1}{2}\sum_{j=1}^{n+1}\left( I_j+\bar I_j \right)\\
=r^2-\frac{1}{2}\sum_{j=1}^{n+1}\left(r\left(\frac{a_j}{w_j}+\frac{\bar a_j}{\bar w_j}\right)+I_j+\bar I_j\right).\end{array}
\label{casimir}
\end{equation}

\begin{proposition} \label{int}

The functions $I_k,\bar I_k, Q$ (\ref{inte}, \ref{casimir}) are linearizing variables of the system (\ref{zzt}) satisfying
\begin{equation}\label{a-a}
\begin{array}{ll}
\partial_{\tau_j} Q  = 0, & \partial _{ { \bar \tau _j } } Q = 0 ,\\
\partial_{\tau_j} I_k=\delta_{jk}, & \partial_{\bar\tau_j} I_k=0 ,\\
\partial_{\tau_j} \bar I_k=0, & \partial_{\bar\tau_j} \bar I_k= - \delta_{jk} . \end{array}
\end{equation}
\end{proposition}
Proof: We first calculate derivatives of each $I_{k}$ with respect
to the times $\tau_j, j=0..n+1$. Applying integration by
parts, we get
$$
2\pi i\frac{\partial I_k}{\partial\tau_j}=\frac{\partial}{\partial\tau_j} \oint_{w_k}\bar z \frac{\partial z}{\partial w} dw= \oint_{w_k}\frac{\partial}{\partial w}\left(\bar z\frac{\partial z}{\partial \tau_j}\right) dw+\oint_{w_k}\left(\frac{\partial\bar z}{\partial \tau_j}\frac{\partial z}{\partial w}-\frac{\partial z}{\partial \tau_j}\frac{\partial \bar z}{\partial w}\right) dw .
$$
By the equations of motion (\ref{zzt}) and the definition of the Lax-Poisson brackets (\ref{Lax bra}) this equals
$$
\oint_{w_k}\frac{\partial}{\partial w}\left(\bar z\frac{\partial z}{\partial \tau_j}\right) dw+\oint_{w_k}\left(\{{\mathcal H}_j,\bar z\}\frac{\partial z}{\partial w}-\{{\mathcal H}_j,z\}\frac{\partial \bar z}{\partial w}\right) dw
$$
$$
=\oint_{w_k}\frac{\partial}{\partial w}\left(\bar z\frac{\partial z}{\partial \tau_j}\right) dw+\oint_{w_k}\frac{\partial {\mathcal H}_j}{\partial w}w\left(\frac{\partial z}{\partial w}\frac{\partial \bar z}{\partial x}-\frac{\partial z}{\partial x}\frac{\partial\bar z}{\partial w}\right)dw ,
$$
which, by the string equation (\ref{string}) reduces to:
$$
\oint_{w_k}\frac{\partial}{\partial w}\left(\bar z\frac{\partial z}{\partial \tau_j}\right) dw  +\oint_{w_k}\frac{\partial {\mathcal H}_j}{\partial w} dw= 2\pi i\delta_{kj}, \quad \delta_{jk}=\left\{\begin{array}{l}1, \quad j=k \\
0,\quad j\not=k .\end{array}\right. 
$$
The first term in the lhs vanishes,
because $\bar z\frac{\partial z}{\partial \tau_j}$ is univalent in a neighbourhood of $w_k, k=1..n+1$ (and at $\infty$ due to (\ref{suma})) and the remaining integral is evaluated by substituting expression (\ref{logh}) for ${\mathcal H}_j$. We have thus proved that
$$
\frac{\partial I_i}{\partial\tau_j}=\delta_{ij} .
$$
The rest of the proposition is proved by similar computations.

\begin{corollary}

The vector fields defining (\ref{zzt}) commute.
\label{logint} 
\end{corollary}

Proof: Since $z$, $\bar z$ are completely determined by the new coordinates $Q,I,\bar I$, infinitesimal deformations of the former under the $\tau_i$ flows can be expressed via the chain rule through the infinitesimal deformations of the latter. It follows from Proposition \ref{int} that
$$
\frac{\delta}{\delta \tau_i}\left(\begin{array}{c} z \\ \bar z \end{array}\right)=\frac{\partial}{\partial I_i}\left(\begin{array}{c} z \\ \bar z \end{array}\right), \quad \frac{\delta}{\delta \bar \tau_i}\left(\begin{array}{c} z \\ \bar z \end{array}\right)=\frac{\partial}{\partial \bar I_i}\left(\begin{array}{c} z \\ \bar z \end{array}\right),
$$
so we have
$$
\left(\frac{\delta}{\delta \tau_i}\frac{\delta}{\delta \tau_j}-\frac{\delta}{\delta \tau_j}\frac{\delta}{\delta \tau_i}\right)\left(\begin{array}{c} z \\ \bar z \end{array}\right)=\left(\frac{\partial}{\partial I_i}\frac{\partial}{\partial I_j}-\frac{\partial}{\partial I_j}\frac{\partial}{\partial I_i}\right)\left(\begin{array}{c} z \\ \bar z \end{array}\right)=0.
$$
The commutativity of the other flows are seen to similarly follow from Proposition \ref{int}.

Choosing the Poisson structure for equations (\ref{a-a}) in a canonical way 
\begin{equation}
\{I_i,\bar I_j\}_p=\delta_{ij}, \quad \{I_i,I_j\} _ p=\{\bar I_i, \bar I_j\} _ p=0, \quad \{ Q , \bar I _j \} _p = \{ Q, I _j \} _p  = 0 ,
\label{true}\end{equation}
we may interpret $I,\bar I$ as canonical linearizing variables and $Q$ as a Casimir invariant.
The equations of motion induced by the evolution operators ${\mathcal H}_i$ or $\bar {\mathcal H}_i$ are seen to be Hamiltonian and generated by the Hamiltonians $\bar I _i, \quad I _i$.

The Poisson brackets $\{ , \}_p$ in
(\ref{true}) are different from $\{ , \}$ in (\ref{Lax bra}); they
define a Poisson structure on the finite dimensional space of logarithmic
reductions of the phase space satisfying the string equation, while the latter (the Lax-Poisson
bracket) is a dispersionless limit of the commutator.

Remark: Linearizing coordinates similar to (\ref{inte}) appeared in  \cite{MD}  in connection with the Laplacian Growth problem as
integrals of the Laplacian Growth
(string) equation, without the introduction of the compatible $\tau_i$ and $\bar \tau_i$ flows considered here.

The following proposition makes explicit the fact that the string equation implies the finite dimensionality of the reduced phase space.

\begin{proposition}
The functions (\ref{inte}) form a set of integrals of the string
equation (\ref{string}). The string equation
(\ref{string}) implies
$$
\frac{\partial I_i}{\partial x} = \frac{\partial \bar I_i}{\partial x}= 0,\quad \frac{\partial Q}{\partial x}= 1.
$$
\label{qi}
\end{proposition}
Equivalently, this may be integrated to
$$
Q-x=c_0, \quad I_i=c_i, \quad \bar I_i=\bar c_i, \quad i=0..n+1,
$$
where $c_i, c_i$ are constants in $x$, which may be interpreted as coordinates on the $2n+5$ dimensional reduced phase space.

Proof: This is similar to the proof of proposition (\ref{int});
one simply differentiates with respect to $x$ and evaluates corresponding residues.

As mentioned in the Introduction, the Laplacian growth problem is recovered by identifying $\bar z$ as the complex conjugate of $z$. As seen from Proposition \ref{qi}, the Casimir $Q$, which is proportional to the area of ${\mathcal D}_+$ grows with unit speed in physical time $x$, while $I_k,\bar I_k$ are functions of the harmonic
moments of the boundary curve.

To obtain similar results for the rational case we take a
limit as in (\ref{ratlim}).

\begin{corollary}
Let $z,\bar z$ be of the rational reduction form (\ref{bar-z-add}).
Then the following $2n+3$ quantities
$$
\begin{array}{ll}
I_0=\bar u_0-\sum_{i=1}^n\bar u_i / \bar w_i, &  \bar I_0=u_0-\sum_{i=1}^n u_i /  w_i \\
I_{2i-1}=rw_i^{-1}+\bar u_0+\sum_{j=1}^n \frac{\bar u_j}{1/w_i - \bar w_j}, & \bar I_{2i-1}=r \bar w_i^{-1}+u_0+\sum_{j=1}^n \frac{u_j}{1/\bar w_i -  w_j}\\
I_{2i}=\left(r-\sum_{j=1}^n \frac{\bar u_j}{(1/w_i - \bar w_j)^2}\right) \frac{u_i}{w_i^2}, & \bar I_{2i}=\left(r-\sum_{j=1}^N \frac{ u_j } { (1/\bar w_i - w_j)^2 } \right)\frac{ \bar u_i} {\bar w_i^2}
\end{array}, \quad i=1..n
$$
$$
Q=r^2-\frac{1}{2}\sum_{i=1}^n\left(r\left(\frac{\bar u_i}{\bar w_i^2}+\frac{u_i}{w_i^2}\right)+\bar I_{2i}+I_{2i}\right)
$$
the linearizing canonical variables for the rational reduction, i.e. variables in terms of which the equations of motion (\ref{eqs_poles}) for the rational reduction (\ref{bar-z-add}) have the canonical form (\ref{a-a}).
\end{corollary}

The logarithmic or rational flows are infinite linear combinations of the 2dToda flows (\ref{zz}). As shown in Appendix B, harmonic moments of the conformal mapping $z(w)$ are linear in the 2dToda times. Therefore to express $t_i, \bar t_i, i=1,2..$ through $\tau_i, \bar \tau_i, i=0..n+1$ one has to evaluate the integrals
\begin{equation}
M_k(z,\bar z)=\frac{1}{2k\pi i}\oint_{\infty}\bar z z^{-k} \frac{\partial z}{\partial w}dw, \quad \bar M_k(z,\bar z)=\frac{1}{2k\pi i}\oint_{0}z \bar z^{-k} \frac{\partial \bar z}{\partial w}dw,
\label{ztolog}
\end{equation}
which are functions of coordinates $z, \bar z$ of the reduced phase space (\ref{zln}),  and then express $z, \bar z$ through $Q,I_i,\bar I_i, i=0..n+1$. Then
\begin{equation}
M_k(Q,I,\bar I)=M_k(z(Q,I,\bar I),\bar z(Q,I,\bar I)), \quad \bar M_k(Q,I,\bar I)=M_k(z(Q,I,\bar I),\bar z(Q,I,\bar I))
\label{logtot}
\end{equation}
Since, modulo integration constants
$$
M_i=t_i, \quad \bar M_i=\bar t_i, \quad i>0 ,
$$
$$
I_i=\tau_i, \quad \bar I_i=\bar \tau_i,\quad i=0..n+1
$$
and $Q=x$, it follows that equation (\ref{logtot}) expresses $t_i, \bar t_i$ in terms of $\tau_i, \bar \tau_i$.

\end{subsection}

\label{poisson}

\end{section}

\begin{section}{Conclusions.}

We have derived consistent finite dimensional logarithmic and rational reductions of certain flows related to the 2dToda hierarchy.  The requirement for the consistency of these reductions was the string equation. Since the latter may also be viewed as the Galin-Polubarinova  equation for the Hele-Shaw (Laplacian growth) process, this also established the latter as a compatible constraint for such reduced 2dToda flows.

More generally, it would be of interest to determine all finite dimensional reductions of the 2dToda hierarchy that imply fulfilment of the Darcy law in various forms.
 
Also, it would be useful to study integrable systems connected to various kinds of Hele-Shaw flows. This could include boundary conditions which are more general than a point sink at infinity. For instance, one can consider an exterior problem with a steady and uniform viscous flow at infinity. This can describe the evolution of a bubble surrounded by a viscous liquid moving through a wide channel, or a bubble within a viscous flow generated by a source and a sink of equal magnitude separated by a large distance. In such models, the constraint on $z, \bar z$ (\ref{z}, \ref{z-bar}) has the form
$$
\{z(w,x), \bar{z}(w^{-1},x)\} = (1/w+w)r(x)
$$ 
This is an exterior analog of the interior problem with a dipole source inside the domain. In contrast to a monopole sink, this implies that the first harmonic moment changes with the time $x$, while the others (including area) remain unchanged. In general any combination of $n$-pole sources could be considered:
$$
\{z(w,x), \bar{z}(w^{-1},x)\} = f(x,w)
$$
where $f$ is defined by the asymptotics of the hydrodynamic potential.

\end{section}

\begin{section}{Appendix A: String equation and Darcy law}

We recall here the derivation of the Galin-Polubarinova equation. Consider evolution of the exterior $ {\mathcal D}_- $ of a simply-connected planar domain bounded by an analytic curve on $z$-plane. The curve is an image of the unit circle in the $w$ plane under a conformal mapping (\ref{un}). As stated in the introduction the pressure field $P(X,Y)$ is constant in the interior domain ${\mathcal D}_+$ and at the boundary $\partial {\mathcal D}_+$, so that at $\partial {\mathcal D}_+$
$$
0=dP/dx=\partial P/\partial x+v_n\nabla P=\partial P/\partial x-(\nabla P)^2,
$$
where we used Darcy law (\ref{normal}) and the pressure gradient is taken in the exterior vicinity of the curve. Since $P$ is harmonic in ${\mathcal D}_-$ (cf (\ref{laplace})), it is the real part of a homomorphic function (hydrodynamic potential)
$P={\rm Re}\phi(z)$. It follows that
$$
0={\rm Re}\left(\frac{\partial \phi}{\partial x} - \frac{\partial \phi}{\partial z}\overline {\left( \frac{\partial \phi}{\partial z} \right)}\right) .
$$
The hydrodynamic potential can be chosen as $\phi(z,x)=1/2\pi\log w(z,x)$ due to the logarithmic asymptotic in (\ref{laplace}). Then
$$
0={\rm Re}\left(\overline { w(z,x)}\frac{\partial w(z,x)}{\partial x}-\frac{\partial w(z,x)}{\partial z}\overline{\left(\frac{\partial w(z,x)}{\partial z}\right)}\right)
$$
where $w(z,x)$ stands for the map inverse to $z(w,x)$. Using the facts that $z=z(w(z,x),x)$; i.e.,
$$
\frac{\partial z(w,x)}{\partial w}\frac{\partial w(z,x)}{\partial x}+\frac{\partial z(w,x)}{\partial x}=0, \quad \frac{\partial w(z,x)}{\partial z}=1/\frac{\partial z(w,x)}{\partial w} ,
$$
and $\bar w=1/w$ on $\partial {\mathcal D}_+$, we arrive at (\ref{inter}).
\end{section}

\begin{section}{Appendix B: Poisson structure of polynomial reductions}

Following steps similar to the proof of Proposition \ref{int}, one arrives at the following result for polynomial reductions
(\ref{poly1}), (\ref{poly2}), (\ref{dflows}) of the 2dToda system.

\begin{proposition}
The following quantities 
$$
M_k=\frac{1}{2i\pi k}\oint_{\infty}\bar z z^{-k} d z, \quad \bar M_i=\frac{1}{2i\pi k}\oint_{0} z \bar z^{-k} d \bar z,
$$
$$
Q=\frac{1}{4 i\pi}\left (\oint_{\infty} \bar z dz - \oint_{0} z d \bar z\right)
$$
are canonical linearizing variables of the (polynomially) reduced 2dToda system (\ref{poly1}), (\ref{poly2}), (\ref{dflows}), (\ref{zz}):
$$
M_k(z,\bar z)=t_k+{\rm const}, \quad \bar M_k(z,\bar z)=\bar t_k+{\rm const}, \quad Q(z,\bar z)=x+{\rm const} .
$$
Solutions to the string equation (\ref{string}) are then defined by the set of algebraic equations
$$
Q(z,\bar z)=x+c_0, \quad  M_k(z,\bar z)=c_k, \quad \bar M_k(z,\bar z)=\bar c_k,
$$
where $c_k, \bar c_k$ are arbitrary constants.
\label{prop}
\end{proposition}

Proof: Evaluating $\frac{\partial M_j}{\partial t_k}, \frac{\partial M_j}{\partial \bar t_k}$ by arguments similarly to those of proposition \ref{int}, we get
$$
\frac{\partial M_j}{\partial t_k}=-\frac{1}{2i\pi j}\oint_{\infty}\frac{\partial H_k}{\partial w}z^{-i}dw, \quad \frac{\partial M_i}{\partial \bar t_j}=\frac{1}{2i\pi j}\oint_{\infty}\frac{\partial\bar H_k}{\partial w}z^{-i}dw .
$$
Using the fact that
$$
H_k=(z(w)^k)_++\frac{1}{2}(z(w)^k)_0=z(w)^k-(z(w)^k)_--\frac{1}{2}(z(w)^k)_0
$$
and the negative part of the Laurent expansion does
not contribute to the integral, we obtain
$$
\frac{\partial M_j}{\partial t_k}=\frac{1}{2i\pi j}\oint_{\infty}z(w)^{-i}\frac{\partial z(w)^j}{\partial w}dw=\delta_{kj} .
$$
Using the fact that the evolution functions $\bar H_j = (\bar z^j)_-+1/2(\bar z^j)_0 $ generating the $\bar\tau_i$ flows, are polynomials of nonegative degrees in $1/w$, we obtain
$$
\frac{\partial M_i}{\partial \bar t_j}=0  .
$$

In the context of the Hele-Shaw problem, the first $N+1$ Richardson moments \cite{R} equal the first $N+1$ 2dToda times. Indeed, in the Hele-Shaw problem $z(w)$ has the meaning of a conformal mapping from $w$
to the $z$ plane. In the exterior problem, the exterior of the unit circle (i.e. domain $|w|> 1 $) is mapped to the exterior ${\mathcal D}_- $ of the of the boundary curve $z(w)$. The exterior harmonic moments are, by Green's theorem
$$
\frac{1}{2j\pi i}\int_{|w|> 1} z^{-j} dz d\bar z =\frac{1}{2j\pi i} \oint_{|w|=1} \bar z(1/w)z(w)^{-j} dz .
$$
Since the mapping $z(w)$ is univalent in ${\mathcal D}_-$, all zeros of $z(w)$ are located in ${\mathcal D}_+$ and we can move the integration contour to infinity if $i\ge1$. Thus, we get
$$
\frac{1}{2j\pi i}\oint_{\infty} \bar z(1/w)z(w)^{-j} dz=M_i
$$
which, by proposition \ref{prop} equals $t_i$ modulo an integration constant.
\end{section}

\begin{section}{Appendix C: Poisson structure of 1dToda hierarchy.}

In this section we consider the Poisson structure of rational reductions of 1dToda system.

Recall that for the 1dToda system one takes into account only the $t_i$ flows.
\begin{equation}
\partial_{t_i}z=\{H_i,z\}, \quad H_i=(z(w)^i)_++1/2(z(w)^i)_0, \quad i=0..\infty
\label{self}
\end{equation}
This system is bi-Hamiltonian (for general information e.g. see \cite{D}) with two (linear and quadratic) compatible Poisson structures. For generic Toda system (\ref{z}), the dispersionless linear Poisson
brackets for the ''field variables"
$u_i, i=1..\infty$ (\ref{z}) have the following form \cite{C}
\begin{equation}
\{u_n(x), u_m(y)\}_1=2(c_n+c_m-1) \big[ (n+m)u_{n+m}(x) \delta '(x-y)+m
u_{n+m}'(x)\delta(x-y) \big] 
\label{lin} \end{equation}
where
\begin{displaymath}
c_k=\left\{ \begin{array}{ll} 1 & \textrm{, if $k>0$}\\
1/2 & \textrm{, if $k=0$}\\
0 & \textrm{, if $k<0$}
\end{array} \right.
\end{displaymath}
while the quadratic brackets are
\begin{eqnarray}
\{u_n(x),u_m(y)\}_2=\big[\frac{1}{2}(n-m)u_n(x)u_m'(x)+(\sum\limits_{k=1}^{1-n}(n-m+k)u_{n+k}(x)u_{m-k}'(x)+\nonumber\\
+ku_{n+k}'(x)u_{m-k}(x))\big]\delta(x-y)+\big[1/2(n-m)u_nu_m+ \nonumber\\
+(\sum\limits_{k-1}^{1-n}(n-m+2k)u_{n+k}u_{m-k})\big]\delta(x-y) .  \label{qua}
\end{eqnarray}

As seen from the lemma \ref{Rat-half-flows}, the rational functions
\begin{equation}
z(w)=\frac {q_{N+1}(w)} {p_{N}(w)} = \frac {w^{N+1}+\sum_{i=0}^{N} a_i w^i } {\sum_{i=0}^{N} b_i w^i}\label{proper}
\end{equation}
are form-invariant under all the 1dToda flows (\ref{self}), without any extra restriction (e.g. no string equation is needed).

We obtain corresponding Poisson structures for the variables $a_i, b_i$, by using result (\ref{qua}), expressing
$u_i$ in terms of $a_i,b_i, i=0..N$. Both the linear (\ref{lin}) and the quadratic (\ref{qua}) Poisson structures lead to quadratic brackets
for $a_i, b_i$ . Namely , the second Poisson structure for (\ref{proper}) becomes
\begin{eqnarray}
\{a_k(x),a_l(y)\}_2=\big[\sum\limits_{n=1}(l+n-k)a_{k-n}(x)a_{l+n}(y)+na_{k-n}(y)a_{l+n}(x))\nonumber\\
+(l-N-1)a_k(x)a_l(y)\big]\delta'(x-y) ,  \label{aa-qua}
\end{eqnarray}
\begin{eqnarray}
\{b_k(x),b_l(y)\}_2=\big[\sum\limits_{n=1}(k-l-n)b_{k-n}(x)b_{l+n}(y)-nb_{k-n}(y)b_{l+n}(x))+\nonumber\\
+\frac{k-l}{2}b_k(x)b_l(y)\big]\delta'(x-y) ,  \label{bb-qua}
\end{eqnarray}
\begin{eqnarray}
\{a_k(x),b_l(y)\}_2=\small{\frac{k-N-1}{2}}a_k(x)b_l(y)\delta'(x-y).  \label{ab-qua}
\end{eqnarray}
The first Poisson structure can be obtained from (\ref{aa-qua} - \ref{ab-qua}) with the help of a shift by a constant
$$
a_i\to a_i+\lambda b_i, \quad z(w,x)\to z(w,x)+\lambda
$$
and using a bi-Hamiltonian nature of (\ref{lin}), (\ref{qua}).
\begin{eqnarray}
\{a_k(x),a_l(y)\}_1=\big[\sum\limits_{n=1}(k-l-n)(a_{k-n}(x)b_{l+n}(y)+b_{k-n}(x)a_{l+n}(y))-\nonumber\\
-n(a_{k-n}(y)b_{l+n}(x)+b_{k-n}(y)a_{l+n}(x)))+\frac{N+1-l}{2}b_k(x)a_l(y)+ \nonumber\\
+\frac{k+N+1-2l}{2}a_k(x)b_l(y)\big]\delta'(x-y),  \label{aa-lin}
\end{eqnarray}
\begin{eqnarray}
\{a_k(x),b_l(y)\}_1=\big[\sum\limits_{n=1}((k-l-n)b_{k-n}(x)b_{l+n}(y)-nb_{k-n}(y)b_{l+n}(x))+\nonumber\\
+\frac{N+1-l}{2}b_k(x)b_l(y) \big]\delta'(x-y),   \label{ab-lin}  \end{eqnarray}
\begin{eqnarray}
\{b_k(x),b_l(y)\}_1=0. \label{bb-lin}
\end{eqnarray}
In all the above expressions $a_{N+1}=1$ and $ a_i=0 $ if $i$ goes beyond the range $i=0..N+1$ (and $b_j=0$ if $j\not=0..N$).

These brackets form a bi-Hamiltonian structure for rational reductions of 1dToda hierarchy:
\begin{equation}
\partial_{t_i}z=\{{\rm H}_i,z\}_1=\{{\rm H}_{i-1},z\}_2
\end{equation}
with the following Hamiltonians:
\begin{equation}
{\rm H}_i=\frac{1}{i+1}\int (z^{i+1}(x))_0 dx
\end{equation}

\end{section}

\vspace{5mm}

{\bf\Large Acknowledgements}\\*[3ex]

The authors would like to acknowledge helpful discussions and usefull information received from B.Dubrovin, S.Howison, M.Mineev, J.Ockendon, A.Orlov and M.Pavlov. The work of O.Y. and I.L. was supported by the European Community IIF's MIF1-CT-2004-007623 and MIF1-CT-2005-007323.

\end{document}